\documentclass[pra,twocolumn,a4paper,superscriptaddress, showpacs]{revtex4}
\usepackage{amsmath}
\usepackage{amsfonts}
\usepackage{graphicx}
\begin{document}
\title {Quantum reading of digital memory with non-Gaussian entangled light}
\author{J. Prabhu Tej}
\affiliation{Department of Physics, Bangalore University, 
Bangalore-560 056, India}
\author{A. R. Usha Devi}
\email{arutth@rediffmail.com}
\affiliation{Department of Physics, Bangalore University, 
Bangalore-560 056, India}
\affiliation{Inspire Institute Inc., Alexandria, Virginia, 22303, USA.}
\author{A. K. Rajagopal}
\affiliation{Inspire Institute Inc., Alexandria, Virginia, 22303, USA.}
\affiliation{Harish-Chandra Research Institute, Chhatnag Road, Jhunsi, Allahabad 211 019, India.}
\begin{abstract}
It has been shown recently (Phys. Rev. Lett. 106, 090504 (2011)) that entangled light with Einstein-Podolsky-Rosen (EPR) correlations  retrieves information from digital memory better than any classical light. In identifying this, a  model of digital memory with each cell consisting of reflecting medium with two  reflectivities (each memory cell encoding the binary numbers 0 or 1) is employed. The readout of  binary memory  essentially corresponds to discrimination of two Bosonic attenuator channels characterized by  different reflectivities. The model requires an entire mathematical paraphernalia of continuous variable  Gaussian setting for its analysis, when arbitrary values of reflectivities $0\leq r_0,r_1\leq 1$ are considered. Here we restrict to a basic quantum read-out mechanism with two different families of non-Gaussian entangled states of light, with the binary channels to be discriminated being (i) ideal memory characterized by reflectivity $r_1= 1$ (identity channel) and (ii) a thermal noise channel --  where the signal light illuminating the memory location gets completely lost ($r_0=0$) and only a white thermal noise hitting the upper side of the memory reaches the decoder. We compare the quantum reading efficiency of two families of non-Gaussian entangled light (${\cal M}\&\, {\cal M}$ family of path entangled photon states and entangled state obtained by mixing single photon with coherent light in 50:50 beam splitter) with any classical source of light in this  model. We identify that the classes of non-Gaussian entangled transmitters studied here offer significantly better reading performance than any classical transmitters of light in the regime of low signal intensity. We also demonstrate that ${\cal M}\&\, {\cal M}$  family of entangled light exhibits improved reading performance than N00N states.

\end{abstract}


\maketitle

\section{Introduction}

Entangled states are known to offer enhanced performance sensitivity over unentangled ones in quantum channel  discrimination~\cite{Sac, piani}. This opens up several potential implications in quantum information protocols. Formulating target detection as a channel discrimination problem,  advantage of entangled light transmitter over  a classical source of light has been explored in Refs.~\cite{Tan, Lloyd, Shapiro, Guha, ARU, Yuen}, where  light received from a far target region is used to ascertain if a low reflectivity object -- immersed in a bright thermal noise -- is present or not. In a different regime, EPR correlated light is shown to offer remarkable improvement in the read-out of information from  a digital memory over any classical light~\cite{pirandola}. Here, a binary memory system  is modelled as an array of cells, where each memory cell is composed of a  reflecting medium with two possible reflectivities  to store a bit of information. The task of digital readout in this model is essentially to discriminate the two channels characterized by beam splitters with reflectivities $r_0$, $r_1$, by distinguishing the states of reflected light. In the basic scheme of reading digital memory, a transmitter emits a global state of light, composed of $M$ copies of signal (with $N_S$ mean number of photons per signal mode) and $L$ copies of idler. The signal light illuminates the memory cell of optical reflectivity $r_0$ (encoding bit value 0) or $r_1$ (encoding bit value 1)  and the reflected light is detected along with the idler at the receiver. A optimal measurement scheme is employed to decode the bit value 0 or 1 with an error probability $P_e$. For a fixed total mean number $N=M\, N_S$ of signal photons illuminating each memory cell, it was shown that a non-classical transmitter emitting EPR correlated light can retrieve more information than any classical source of light, in the regime of few photons ($N<10^2$) and for typical optical memories with high reflectivities~\cite{pirandola}. Following Ref.~\cite{pirandola}, the use of non-classical  transmitters of light to read  digital memory has been referred to as {\em quantum reading}. Several other supplementary features of quantum reading have been explored recently~\cite{Nair, pirandola2, Hirota, Guha2, Bisio, MM, Dall, braunstein, Dall2, Dall3}.    

In this paper we consider a simple model of digital memory consisting of cells of reflectivity $r_1=1$, encoding bit value 1 (corresponds to the situation where signal light is  reflected without any loss)  and $r_0=0$, encoding bit value 0 (corresponds to total loss of signal light, while a weak thermal noise is received by the detector). The strategy of readout thus reduces to the discrimination of identity and thermal channels i.e.,  reading memory in this case corresponds to distinguishing  input  light  and a weak thermal light. If the input light is chosen to be a pure state, this scheme leads to a simple analysis to obtain analytical results for various bounds on error probability. This motivates us to explore the quantum advantage of two different classes of non-Gaussian entangled pure input states of light  vs any classical light in digital memory reading. 

It may be noted that in the original scheme Pirandola~\cite{pirandola} had shown that the two mode squeezed vacuum (TMSV) light is more efficient in reading digital memory than any classical light, in the regime of few photons shining the optical memory cells of high reflectivities. This inspired further studies where non-Gaussian, non-classical sources of light like Fock states~\cite{Nair, pirandola2} and path entangled N00N states~\cite{pirandola2} are shown to be advantageous over any classical light and they  outperform TMSV too in the faint light regime for some specific reflectivities.  Exploring quantum reading possibilities with various other non-classical sources of light and identifying the regimes where enhancement over classical light is of significance. Here, we study the  reading efficiency of (i) a class  of path entangled bipartite states of photons (known popularly as ${\cal M}\& \,{\cal M}$ states), proposed in Ref.~\cite{Dowling} in the context of robust quantum optical metrology and (ii)  entangled states of light produced by combining a single photon with a  coherent state in a 50:50 beam-splitter~\cite{Gisin}. We carryout a comparison of quantum-classical read-out performances under a {\em local energy constraint}~\cite{braunstein} i.e., by fixing the mean number $N_S$ of photons per signal mode illuminating the memory cell and the number $M$ of signal modes -- in contrast to the analysis with a {\em global energy constraint}~\cite{pirandola}, where the  total average number of photons $N=M\, N_S$ shining the memory is held fixed. Both these transmitters are recognized to be advantageous compared to any classical transmitters in the low signal intensity regime. Our investigation also reveals that the ${\cal M}\& \,{\cal M}$ family of entangled light exhibits improved reading performance than N00N states.  

We organize the  paper  as follows: In Sec.~II we outline preliminary concepts of quantum reading in general. We describe our basic model of memory where  bit value 1 is encoded in a cell consisting of a perfect mirror of reflectivity  $r_1=1$ -- returning the light to the receiver without any loss --  and bit value zero is encoded in a cell of reflectivity zero -- sending thermal light to the  receiver. We illustrate in Sec.~III, the quantum advantage in the readout of  classical information stored in the digital memory, which can be acheived by using non-Gaussian entangled light, in the regime of low signal intensity. Section IV has concluding remarks.                       

\section{Basic model of reading digital optical memory}   

In the model proposed by Pirandola~\cite{pirandola},  storage of binary data 0, 1 in a digital memory corresponds to  encoding them in channels (cells of the  memory) ${\cal E}_{0},\ {\cal E}_{1}$,  which are beam splitters of reflectivities $r_0$ and $r_1$ respectively.  Readout is a process of channel decoding, which corresponds  to discriminating the channels. This is carried out by sending  input  light to illuminate the cells and then distinguishing the reflected output states of light with the help of suitable measurements at the receiving end. 

Let us consider an input Bosonic density matrix $[\rho_{\rm in}]^{\otimes (M,M')}$ consisting of $M$ copies of signal (S) and $M'$ copies of idler (I),  each copy of the  signal mode carrying mean photon number $N_S$. The output state of light $[\rho^{(u)}_{\rm out}]^{\otimes(M, M')}$ received at the detector is a combined state  of  $M$ signal modes  reflected after illuminating a cell of reflectivity $r_u,\ u=0,1$ and $M'$ idler modes: 
\begin{equation}
[\rho^{(u)}_{\rm out}]^{\otimes(M, M')}=\left({\cal E}_u^{\otimes M}\otimes {\cal I}^{\otimes M'}\right)\left[\rho_{\rm in}^{\otimes(M,M')}\right],         
\end{equation}
where the channel ${\cal E}_u^{\otimes M}$ acts on $M$ signal modes and the identity channel ${\cal I}^{\otimes M'}$ operates on $M'$ idler modes.   
Minimum error probability in discriminating the two states $[\rho^{(u)}_{\rm out}]^{\otimes(M, M')}, \ u=0,1$ for a given input state $[\rho_{\rm in}]^{\otimes(M, M')}$  is given by~\cite{Hel, Hol}   
\begin{equation}
\label{err}
P_{\rm err}=\frac{1}{2}\left[1-\frac{1}{2}\left\vert\left\vert [\rho^{(0)}_{\rm out}]^{\otimes(M, M')}-[\rho^{(1)}_{\rm out}]^{\otimes(M, M')}\right\vert\right\vert\right],
\end{equation}
with  $\vert\vert A\vert\vert={\rm Tr}[\sqrt{A^\dag\, A}]$ denoting the tracenorm of $A$, which is the sum of absolute eigenvalues of $A$. 

Average information, in bits, retrieved from digital memory  is quantified by~\cite{pirandola} 
\begin{equation}
J=1-H(P_{\rm err})
\end{equation}     
where $H(x)=-x\log_2\,  x-(1-x)\, \log_2 (1-x)$ is the binary Shannon entropy.  

Finding the eigenvalues of $[\rho^{(0)}_{\rm out}]^{\otimes(M, M')}~-~[\rho^{(1)}_{\rm out}]^{\otimes{(M, M')}}$, in order to evaluate the error probability (\ref{err}), is a  hard computational problem. However, several easier-to-compute upper and lower bounds on error probability  are found useful~\cite{kargin, acin, PirLloyd}. Of interest is the quantum Chernoff bound~\cite{acin}, which gives the  asymptotic rate exponent of error probability with $M$ signal modes as, 
\begin{equation}
\label{peqn1}
P_{\rm err}\leq P_{{\rm err},{\rm QCB}}:=\frac{1}{2}\,\left(\inf_{0\leq s\leq 1}\, {\rm Tr}\{
[\rho^{(0)}_{\rm out}]^s[\rho^{(1)}_{\rm out}]^{1-s}\}\right)^M
\end{equation}

To explore the advantage of non-classical light over any classical light, one needs to check if the probability of error (\ref{err}) with any non-classical light is smaller than that when classical light is employed in the readout process. 

Any Bosonic density matrix $[\rho^{(\rm cl)}]^{\otimes(M,M')}$ of electromagnetic radiation (with $M$ signal modes and $M'$ idler modes)  is classical, when its decomposition in terms of coherent states is positive~\cite{Sudarshan, Glauber}:     
\begin{widetext}
\begin{equation} 
[\rho^{(\rm cl)}]^{\otimes(M,M')}=\int\,  d^2  \alpha_1 \,\ldots \int\,  d^2  \alpha_M\, \int\,  d^2  \alpha'_1 \,\ldots \int\,  d^2  \alpha'_{M'}  P(\alpha_1,\ldots, \alpha_M; \alpha'_1,\ldots, \alpha'_{M'}) 
\otimes_{k=1}^{M}\vert \alpha_k\rangle\langle \alpha_k\vert \otimes_{l=1}^{M'}\vert \alpha'_l\rangle\langle \alpha'_l\vert,
\end{equation}  
\end{widetext}  
where $\{\vert \alpha_k\rangle, (\vert \alpha'_l\rangle)\}$ denote coherent states of the signal (idler) modes; the function $P(\alpha_1,\ldots, \alpha_M; \alpha'_1,\ldots, \alpha'_{M'})$  is  a legitimate probability distribution, being positive and normalized to unity.  It was shown in Ref.~\cite{pirandola} that probability of error with any classical state $[\rho^{(\rm cl)}]^{\otimes(M,M')}$ characterized by a positive ${\cal P}$-representation is lower bounded by a quantity $C(M, N_S, r_0, r_1)$,  which depends  on the number of signal modes $M$, average signal intensity per mode $N_S$, and the reflectivities $r_0$, $r_1$ as,   
\begin{equation}
P_{\rm err}^{\cal C}\geq   C(M, N_S, r_0, r_1):=\frac{1-\sqrt{1-e^{-M\,N_S(\sqrt{r_1}-\sqrt{r_0})^2}}}{2}.
\end{equation}     
It follows that the {\em maximum} information  decoded from digital memory (of reflectivities $r_0,\, r_1$), by employing any classical source of light (with mean intensity $N=M\, N_S$ shining each cell) is given by,    
\begin{equation}
\label{clImax}
J_{{\rm max},{\cal C}}=1-H[C(M, N_S, r_0, r_1)].
\end{equation}
On the other hand, employing a non-classical source of light of  mean intensity $N=M\, N_S$, it is possible to retrieve at least 
\begin{equation}
J_{{\rm min},\, {\cal Q}}=1-H(P_{{\rm err}, {\rm QCB}}) 
\end{equation}
bits of average information. Improvement of non-classical source of light  over any classical transmitter in the readout of information from digital memory is registered if the {\em minimum information gain}~\cite{pirandola} 
\begin{equation}
\label{g}
G(M, N_S, r_0, r_1)=J_{{\rm min},\, {\cal Q}}-J_{{\rm max},{\cal C}}  
\end{equation}     
is positive. It was shown in Ref.~\cite{pirandola} that in the regime of few photons ($N<10^2$) and high reflectivities, non-classical EPR correlated light is capable of  retrieving more information (via positive gain $G$) than any classical light of same average intensity $N=M\, N_S$. In particular,  with small number of signal photons employed to decode {\em ideal memories} $r_1=1$ and $0\leq r_0\leq 1$, it was identified~\cite{pirandola} that the value of $G$ can be remarkably large. 

The quantum effects ($G>0$) persist even when stray thermal background photons (with $N_B\approx 10^{-2}$ --  $10^{-5}$ average number of noise photons per mode) hit the rear side of the memory and reach the receiver along with the reflected signal light~\cite{pirandola, braunstein}. In this case, the  minimum error probability in decoding the memory by any classical light is lower bounded~\cite{braunstein} by a quantity ${\cal C}(M, N_S, N_B, r_0, r_1)$ as 
\begin{eqnarray}
\label{clLB}
P_{{\rm err,\, \cal C}}&\geq &  {\cal C}(M, N_S, N_B, r_0, r_1)\nonumber \\ 
&&:=\frac{1-\sqrt{1-{\cal F}^M(N_S, N_B, r_0, r_1)}}{2}
\end{eqnarray}       
where 
\begin{eqnarray}
{\cal F}(N_S, N_B, r_0, r_1)&=&\frac{{\rm exp}\left[-\frac{(\sqrt{r_0}-\sqrt{r_1})^2}{\gamma}\, N_S\right]}{\sqrt{\gamma^2+\theta}-\sqrt{\theta}}, \nonumber \\
 \gamma=1+(2-r_0&-&r_1)\, N_B,\nonumber  \\
  \theta=4\, N_B^2\, \prod_{u=0,1}(1&-&r_u)[1+(1-r_u)\,N_B]. 
\end{eqnarray}
Substituting ${\cal C}(M, N_S, N_B, r_0, r_1)$ in (\ref{clImax}), and re-expressing (\ref{g}), it is found that for given values of $M, N_S, N_B, r_0, r_1$, a non-classical transmitter emitting EPR correlated light can be used to beat the classical readout (verified via  $G(M, N_S, N_B, r_0, r_1)>0$) even in the presence of thermal noise~\cite{pirandola, braunstein}. In Ref.~\cite{pirandola} the outperformance of non-classical transmitter over classical one was analyzed under a  {\em global energy constraint}, where total average number of photons $N=M\, N_S$ illuminating each memory cell is held fixed. On the other hand, a different kind of quantum-classical comparison is performed based on {\em local energy constraint} i.e., by fixing the mean number of signal photons $N_S$ in each mode and the number of signal modes $M$. 

Here, we confine ourselves to the case of {\em ideal memory} with optical reflectivity $r_1=1$ (${\cal E}_1$ being the identity channel) and $r_0=0$, which corresponds to complete loss of signal photons illuminating the cell, while only stray  thermal photons hitting the upper side of the memory reach the detector (i.e., ${\cal E}_0$ is a thermal channel). This basic model of memory is useful to explore reading abilities of some interesting classes of entangled non-Gaussian  states of light in comparison with classical light.          
     
\section{Quantum reading performance of entangled non-Gaussian states of light}

In this section we explore the quantum efficiency in reading a digital memory with reflectivities $r_1=1$, $r_0=0$ and thermal noise $N_B$, by employing two different families of non-Gaussian entangled states.   

\subsection{Path entangled ${\cal M}\&{\cal M}$ photon  states}
 
Huver et. al.~\cite{Dowling} had proposed a class of bipartite path entangled photon Fock states (which they referred to as ${\cal M}\&\, {\cal M}$ states)  and showed that they perform better than N00N states in the limit of practical quantum optical metrology with appreciable photon loss. N00N states are explored for their  quantum reading advantage  over classical light recently~\cite{pirandola2} and it was shown that when average signal photon intensity $N_S=1$, N00N states perform  better than even TMSV light.  This provides a motivation to examine the more general ${\cal M}\&\, {\cal M}$ class of states (which contain N00N states as a special case) in the read-out of digital memory.      

The family of  two mode photon states, referred to as ${\cal M}\&\, {\cal M}$ states, are  defined as follows: 
\begin{equation} 
\label{mm}
\vert m::m'\rangle = \frac{1}{\sqrt{2}}\, [\vert m, m'\rangle+ \vert m', m\rangle],\ 
\end{equation}       
where $\ m,m'=0,1,2,\ldots $ and $m>m'$. When $m'=0$, the states reduce to the family of N00N states.   

Our focus here is to study the reading performance of this family of photon states in comparison with any classical light. Let us consider $M$ copies of the state with the first mode being signal (S) and the second being idler (I) mode. The average intensity in each signal mode is given by~\cite{note} $N_S=\langle a_S^\dag\, a_S\rangle=(m+m')/2$ for the set of all ${\cal M}\&\, {\cal M}$ states with a given $m$ and $m'$. With $M$ identical copies of the state (\ref{mm}) as input, the corresponding output states of the beam splitter channels ${\cal E}_u,\ u=0,1$, characterized by reflectivities $r_0=0,\ r_1=1$ and thermal bath with $N_B$ noise photons  are given by, 
\begin{widetext}
\begin{eqnarray}
\label{zero}
\ [\rho_{\rm out}^{(0)}]^{\otimes(M,M)}&=& [\rho_{\rm Th}(N_B)]^{\otimes M} \otimes  
\left\{{\rm Tr}_S[\vert m::m'\rangle_{SI}\langle m::m'\vert]\right\}^{\otimes M}\nonumber \\  
&=& [\rho_{\rm Th}(N_B)]^{\otimes M} \otimes  \left\{\frac{1}{2}[\vert m\rangle_{I}\langle m\vert+\vert m'\rangle_{I}\langle m'\vert]\right\}^{\otimes M}, \\
\label{one}
\ [\rho_{\rm out}^{(1)}]^{\otimes(M,M)}&=&[\vert m::m'\rangle_{SI}\langle m::m'\vert]^{\otimes(M,M)} 
\end{eqnarray} 
\begin{figure}
\includegraphics*[width=2.5in,keepaspectratio]{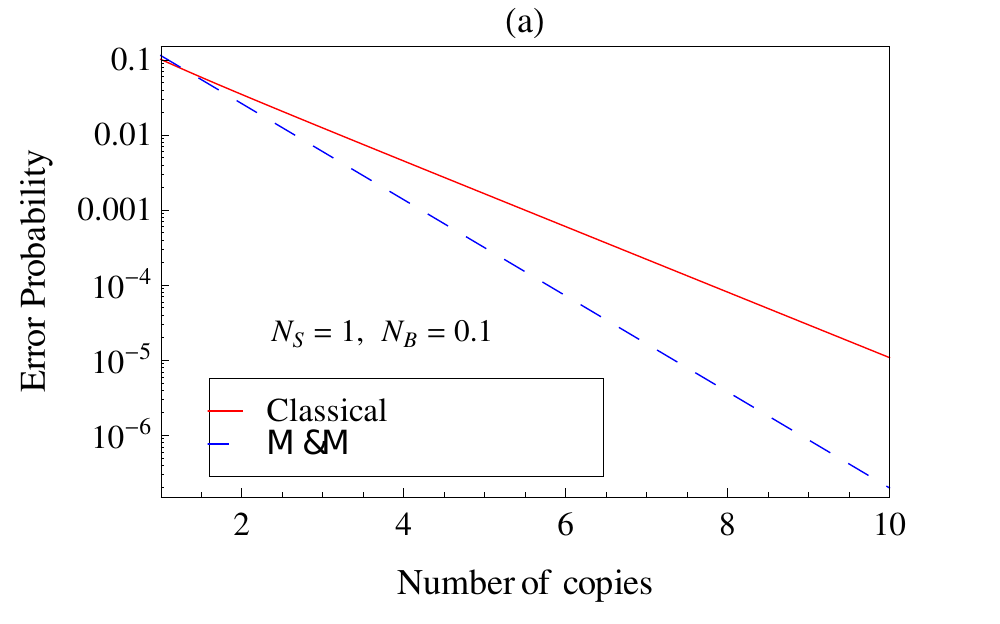}
\includegraphics*[width=2.5in,keepaspectratio]{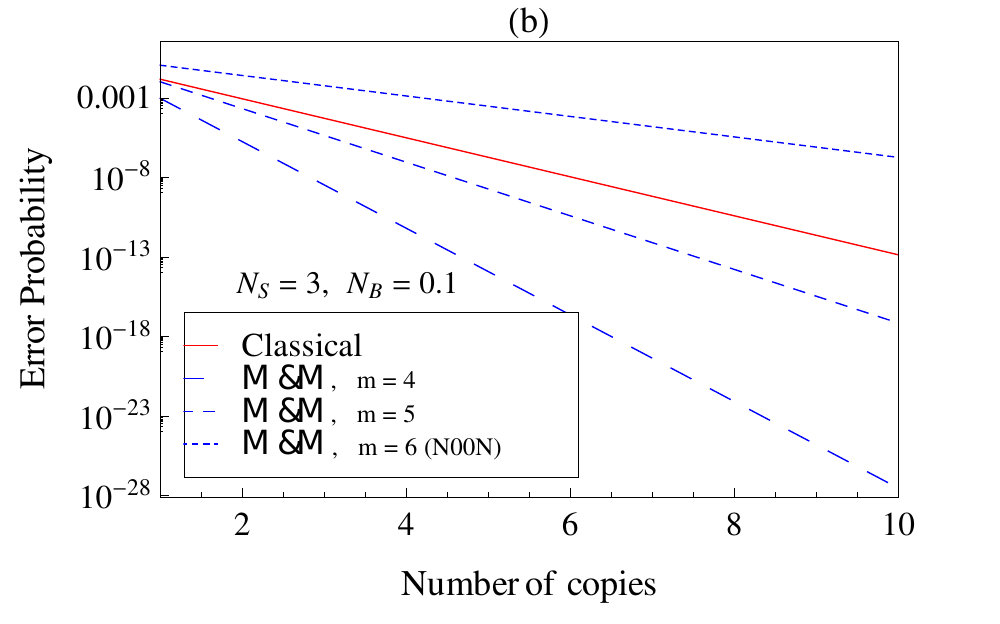}
\caption{(Color online) Quantum Chernoff bound on error $P^{({\cal M}\&{\cal M})}_{{\rm err},{\rm QCB}}(M, N_S, N_B, m)$ of a ${\cal M\& \cal M}$ family of states $\{\vert m::\, 2N_S-m\rangle,\ N_S < m\leq 2N_S\}$  and the classical error bound ${\cal C}(M, N_S, N_B)$ vs number of copies $M$ for the values (a) $N_S=1,\ m=2$  (b) $N_S=3,\ m=4,$  $5, 6$  and thermal noise $N_B=10^{-1}.$ It is clear from Fig. 1b that N00N states ($m=6$) fail to beat  the general ${\cal M}\&\, {\cal M}$ class of states as well as classical light.} 
\end{figure}

\begin{figure}
\includegraphics*[width=2.5in,keepaspectratio]{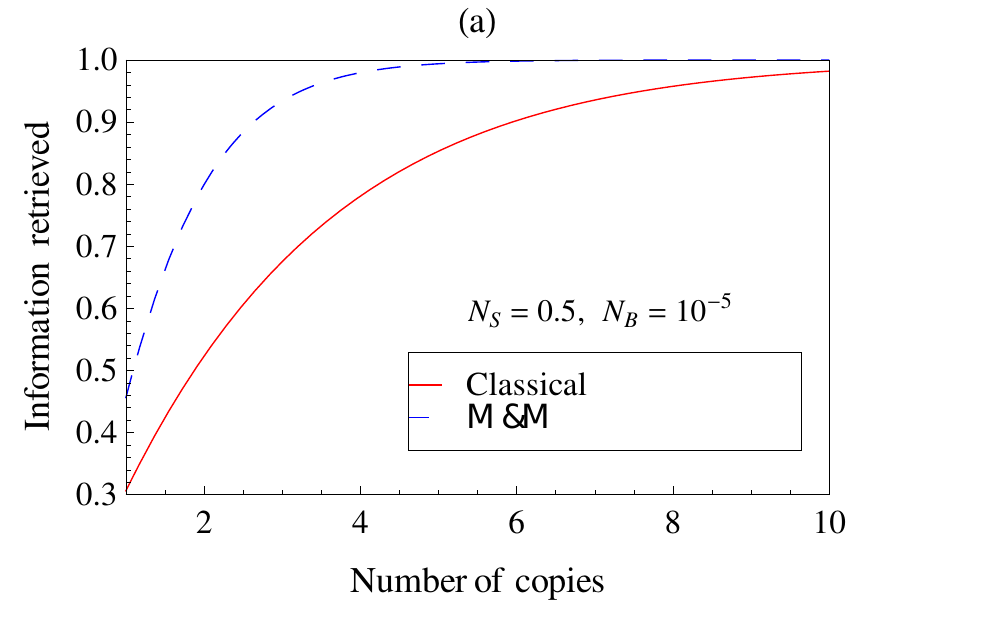}\includegraphics*[width=2.5in,keepaspectratio]{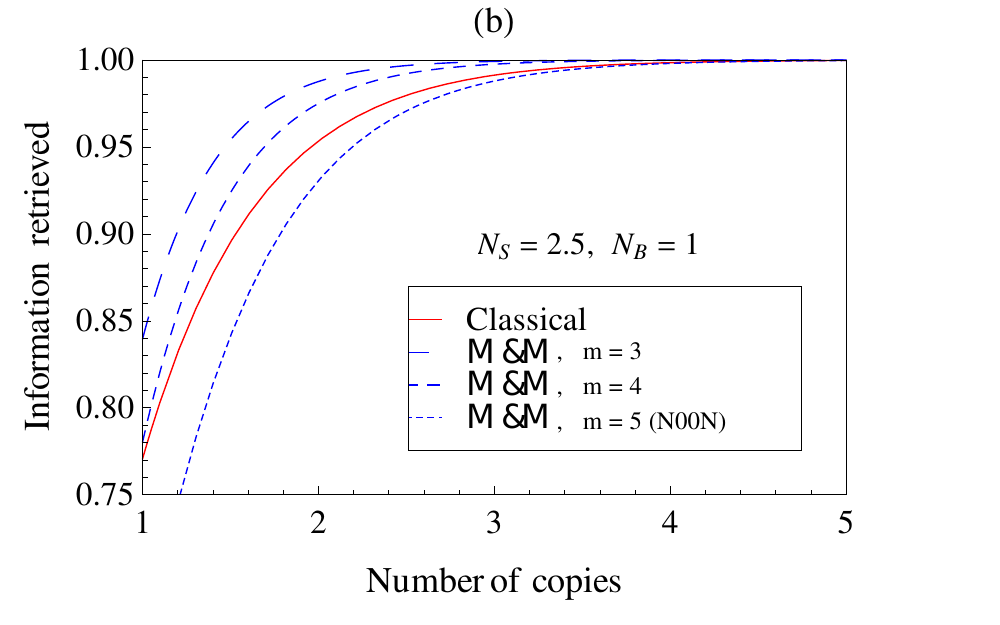}
\caption{(Color online) Maximum information retrieved  $J_{{\rm max},{\cal C}}$ in the readout process from a classical transmitter  and the minimum information decoded from quantum light of ${\cal M}\&{\cal M}$ state for  (a) $N_S=0.5$ and $N_B=10^{-5}$ (b) $N_S=2.5$ and $N_B=1$ with $m=3, 4$ and $m=5$ (N00N state of the family)  plotted as a function of number of copies $M$. It may be seen that quantum reading information retreived via N00N states is smaller compared to that of other states of the ${\cal M}\&{\cal M}$ family and also, classical light.} 
\end{figure}
\end{widetext}
where $\rho_{\rm Th}(N_B)=\displaystyle\sum_{n=0}^{\infty}\frac{N^n_B}{(N_B+1)^{n+1}}\vert n\rangle \langle n\vert$ denotes thermal state of $N_B$ average photons.

As one of the states to be discriminated is pure,  the Chernoff error bound $P_{{\rm err}, {\rm QCB}}$ (see (\ref{peqn1})) for discriminating the output states  (\ref{one}) and (\ref{zero}) can be  readily evaluated (as optimum value of $s$ in (\ref{peqn1}) approaches the value 1 when one of the states is pure and the quantum Chernoff bound reduces to fidelity~\cite{Jphys}): 
\begin{widetext}
\begin{eqnarray}
\label{mmQCB}
P^{{(\cal M}\&{\cal M})}_{{\rm err}, {\rm QCB}}(M, N_S, N_B, m)&=&\frac{1}{2}\, \left(\, _{SI}\langle m::m'\vert\, \left[  \rho_{\rm Th}(N_B)] \otimes  
\left\{\frac{1}{2}[\vert m\rangle_{I}\langle m\vert+\vert m'\rangle_{I}\langle m'\vert]\right\}\right]\vert m::m'\rangle_{SI} \right)^M \\
&=& \frac{1}{2}\,\,\frac{1}{4^M}\, \left(\frac{N_B^m}{(1+N_B)^{m+1}}+\frac{N_B^{m'}}{(1+N_B)^{m'+1}}   \right)^M,\ \ \  N_S=\frac{m+m'}{2}.\nonumber
\end{eqnarray} 
\end{widetext} 
Improved reading efficiency of the ${\cal M}\&{\cal M}$ family of states $\{\vert m::2N_S-m\rangle, N_S< m< 2N_S\}$ over any classical transmitter of the same signal profile $\{M, N_S\}$ is ensured if   the Chernoff bound on error probability (\ref{mmQCB}) -- which affects the readout  --  is smaller than the lower bound (\ref{clLB}) on classical error probability itself i.e.,  $P^{{(\cal M}\&{\cal M})}_{{\rm err}, {\rm QCB}}(M, N_S, N_B, m) < {\cal C}(M, N_S, N_B)$. In Fig.~1 we plot the bounds $P^{{(\cal M}\&{\cal M})}_{{\rm err},{\rm QCB}}(M, N_S, N_B, m),\  {\cal C}(M, N_S, N_B)$ on error probabilities   in logarithmic scale vs the number of copies $M$, for fixed signal energy (a) $N_S=1, m=2$ and (b) $N_S=3,\, m=4, 5, 6$  respectively, with a thermal noise of $N_B=10^{-1}$ photons. We identify that the Chernoff bound on error $P^{{(\cal M}\&{\cal M})}_{{\rm err},{\rm QCB}}(M, N_S, N_B)$ of ${\cal M}\& {\cal M}$ transmitter for $m\neq 2\, N_S$ is smaller than the lower bound on error probability  ${\cal C}(M, N_S, N_B)$  on all classical transmitters, even though both quantum and classical error probability bounds approach zero very sharply with few copies $M$ -- indicating efficient reading possibilities in both situations, for memory cell reflectivities $r_1=1, r_0=0$. However, for $m=2\, N_S, N_S\geq 1.5$, which corresponds to N00N states, the  error probability Chernoff bound is larger than the lower bound on classical transmitters. In other words, N00N states fail to offer quantum reading advantage over classical light when the signal intensity $N_S\geq 1.5$; all other  ${\cal M}\& {\cal M}$ states of the family with a given signal intensity $N_S$  beat the reading performance of N00N states.

These features are also reflected in Fig.~2, where we have plotted the maximum information $J_{{\rm max}{\cal C}}$, retrieved in the classical scenario  and  minimum information $J_{{\rm min},{\cal Q}}$ decoded with quantum ${\cal M}\&{\cal M}$ transmitter in the readout for (a) $N_S=0.5,\, m=1$ and $N_B=10^{-5}$, (b) $N_S=2.5,\, m=3,4,5$ and $N_B=1$. Quantum advantage is evident as  $J_{{\rm max},{\cal C}}<\ J_{{\rm min},{\cal Q}}$ with a single copy   itself for $m=3,4$. However, for $m=5$ (which corresponds to a N00N state) this feature gets reversed i.e., $J_{{\rm max},{\cal C}}>\ J_{{\rm min},{\cal Q}}$. In other words, with increasing signal intensity $N_S$ it is seen that N00N states fail to perform better than classical light, whereas  ${\cal M}\&{\cal M}$ states (excluding the N00N states) perform  significantly better.  It is seen that  both $J_{{\rm max},{\cal C}}$, $J_{{\rm min},{\cal Q}}\rightarrow 1$ with very few number of copies  -- implying perfect readout by both classical and quantum transmitters of light. A positive information gain $G\approx 0.3$ can be achieved for low intensity signals $N_S\leq 1$, even with a single copy of the ${\cal M}\& {\cal M}$ state.  

It is important to point out here that Nair~\cite{Nair} and Pirandola et. al.,~\cite{pirandola2} have shown that Fock state transmitters can outperform the classical transmitters and also entangled TMSV transmitters in the low photon regime for some chosen optical reflectivities. This brings forth a broader question: Is non-classicality of light itself  a resource in memory reading or would it be beneficial to employ entangled light obtained by  mixing non-classical light with classical light in a 50:50 beam-splitter? In the next section we address this question by investigating the reading efficiency of entangled light obtained when single photon Fock state is mixed with  coherent light.

\subsection{Entangled light produced by combining a single photon with coherent light in a 50:50 beam-splitter}

It is wellknown that the output Bosonic state from a 50:50 beam splitter is entangled if  one of the ports has  non-classical state as its input~\cite{Wang, Asboth, Soloman}. Robustness of entanglment of the non-Gaussian  state of light created by combining a single photon  with  coherent light in a beam splitter is analyzed very recently~\cite{Gisin} at various scales by tuning the intensity of the coherent light. We carry out the classical-quantum comparison by studying the reading performance of this family of entangled states.

With a single photon state $\vert 1\rangle$ and coherent state $\vert \alpha\rangle$  of intensity  $\vert\alpha\vert^2$ are sent through a 50:50 beam splitter, the output state (one of the output modes is denoted as signal and the other as idler) is given by, 
\begin{eqnarray}
\label{newstate}
\vert \Psi_{SI}\rangle\ =\ \ {\rm U_{BS}}\, [\, a_S^{\dag}\,\vert 0\rangle _S\otimes D_I(\alpha)\,\vert 0\rangle_{I}\, ] & \\
 =  [D_S(\alpha/\sqrt{2})\otimes D_I(\alpha/\sqrt{2})] &\left[\frac{\vert 1\rangle_S\otimes \vert 0\rangle_I + \vert 0\rangle_S\otimes \vert 1\rangle_I}{\sqrt{2}}\right]. \nonumber  
\end{eqnarray}
Here, the displacement operator $D(\alpha)=e^{\alpha\, a^\dag-\alpha^*\, a}$  generates the coherent state $\vert\alpha\rangle=D(\alpha)\vert 0\rangle$;\  $a_S, a_I$ denote the annihilation operators of the signal and idler modes and the 50:50 beam splitter leads to the transformation $(a_S,\, a_I)\, \stackrel{U_{BS}}{\longrightarrow}\, \left( (a_S-a_I)/\sqrt{2},\  (a_S+a_I)/\sqrt{2} \right)$. The average intensity in the signal mode of the state (\ref{newstate}) is given by $N_S=\langle a_S^\dag\, a_S\rangle=(\vert\alpha\vert^2+1)/2$.

Considering $M$ identical copies of the state (\ref{newstate}), and sending the signal mode to shine the memory cells, the states  of the reflected light combined with the idler modes are given by, 
\begin{widetext}
\begin{eqnarray}
\label{zero_2}
\ [\rho_{\rm out}^{(0)}]^{\otimes(M,M)}&=& [\rho_{\rm Th}(N_B)]^{\otimes M} \otimes  
\left\{{\rm Tr}_S[\vert \Psi\rangle_{SI}\langle \Psi_{SI}\vert]\right\}^{\otimes M}\nonumber \\  
&=& [\rho_{\rm Th}(N_B)]^{\otimes M} \otimes  \left\{D_I\left(\frac{\alpha}{\sqrt{2}}\right)\left[\frac{1}{2}\left(\vert 0\rangle_{I}\langle 0\vert\, +\, \vert 1\rangle_{I}\langle 1\vert\right)\right]D^\dag_I\left(\frac{\alpha}{\sqrt{2}}\right)\right\}^{\otimes M},\\
\label{one_2}
\ [\rho_{\rm out}^{(1)}]^{\otimes(M,M)}&=&[\vert \Psi_{SI}\rangle\langle \Psi_{SI}\vert]^{\otimes(M,M)} 
\end{eqnarray} 
\end{widetext} 
The  Chernoff upper bound on the error probability $P_{{\rm err}, {\rm QCB}}$ affecting the decoding of the binary memory (by discriminating the output states  (\ref{one_2}) and (\ref{zero_2})) is obtained, after simplifications as, 
\begin{widetext}
\begin{eqnarray} 
\label{psiQCB}
P^{(\Psi)}_{{\rm err}, {\rm QCB}}(M, N_S, N_B)&=&\frac{1}{2}\, \left(\, \langle \Psi_{SI}\vert\, \left[  \rho_{\rm Th}(N_B)] \otimes  
\left\{D_I\left(\frac{\alpha}{\sqrt{2}}\right)\left[\frac{1}{2}\left(\vert 0\rangle_{I}\langle 0\vert\, +\, \vert 1\rangle_{I}\langle 1\vert\right)\right]D^\dag_I\left(\frac{\alpha}{\sqrt{2}}\right)\right\}\right]\vert \Psi_{SI}\rangle\right)^M \nonumber \\ 
&=& \frac{1}{2}\, \frac{e^{-\frac{M\,(2N_S-1)}{2(N_B+1)}}}{[4(N_B+1)]^M}\,\left[1+\frac{(2N_S-1)(1+N_B+N_B^2)}{2(N_B+1)^2}\right]^M\,   
\end{eqnarray} 
\end{widetext}

We now carry out the quantum-classical comparison  in reading efficiency of the entangled light $\vert\Psi_{SI}\rangle$  over any classical radiation of the same signal profile $\{M, N_S\}$. In Fig.~3a and 3b it is displayed that the Chernoff bound on error probability  (\ref{psiQCB})  is smaller than the lower bound (\ref{clLB}) on classical error probability (in logarithimic scale), for some typical values of low signal intensities $N_S=1,\  5$, and thermal noise of $N_B=10^{-2}, 1.5$ photons respectively. When we restrict to  average signal intensity $N_S=1$, we can also compare the error probability bounds of the entangled state with the error probability of single photon Fock state~\cite{note1}. We identify that (see Fig.~3a) non-classical single photon Fock state can beat entangled light obtained by mixing it with coherent light of intensity $\vert\alpha\vert^2=1$~\cite{note2}. 

%\begin{widetext}
\begin{figure}
\includegraphics*[width=2.5in,keepaspectratio]{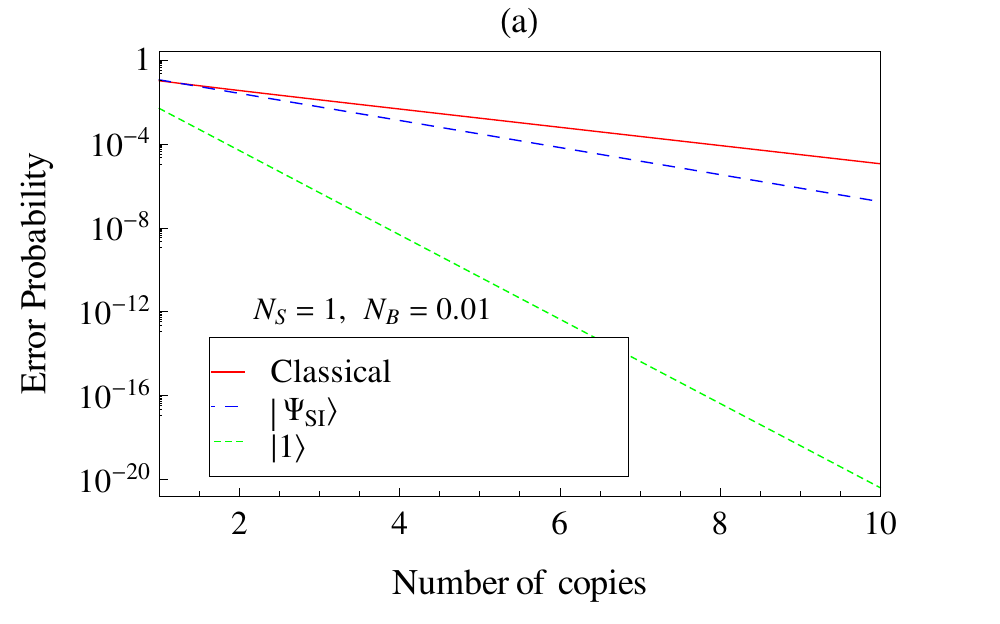}
\includegraphics*[width=2.5in,keepaspectratio]{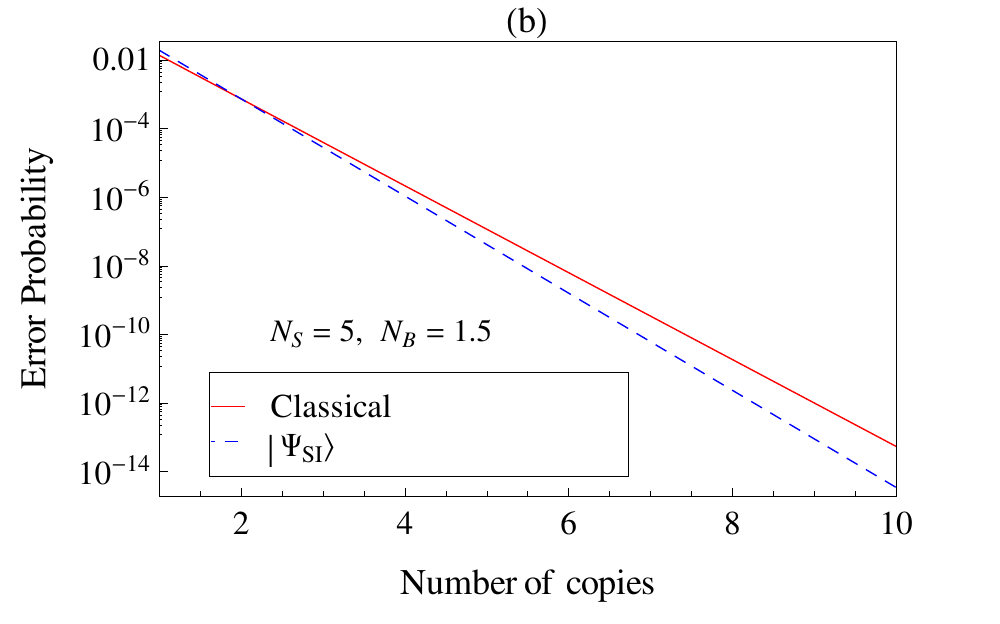}
\caption{(Color online) Error probability bounds $P_{{\rm err},{\rm QCB}}^{(\Psi)}(M, N_S, N_B)$ and ${\cal C}(M, N_S, N_B)$ of the entangled state $\vert \Psi_{SI}\rangle$ (see Eq.~(\ref{newstate})) as a function of number of copies $M$ for different choices of mean signal intensity and thermal noise (a) $N_S=1$, $N_B=0.01$  (b) $N_S=5,\ N_B=1.5$. In Fig.~1a we also plot the error probability of single photon Fock state, which is found to be much smaller than the Cheronoff error bound of the entangled state $\vert \Psi_{SI}\rangle$.}    
\end{figure}
%\end{widetext}

In Fig.~4, we have plotted the maximum decoded information $J_{{\rm max},{\cal C}}$, when the readout process employs classical light  and  minimum information $J_{{\rm min},{\cal Q}}$ that can be retrieved using quantum state (\ref{newstate}) of light. We find that both the informations $J_{{\rm max},{\cal C}}$, $J_{{\rm min},{\cal Q}}\rightarrow 1$  with small number of copies $M$ indicating that with ideal memory $r_1=1$ encoding bit value 1 and  thermal noise mixed with perfectly transmitting  memory $r_0=0$ encoding the bit value 0, the reading efficiency of both classical and quantum light is significantly large. However, there is a quantum advantage and a positive gain of around $G\approx 0.3$ can be realized for average signal photons $0.5\leq N_S\leq 1$ (low intensity regime) with even one copy of the state (\ref{newstate}).

\begin{figure}[h]
\includegraphics*[width=2.5in,keepaspectratio]{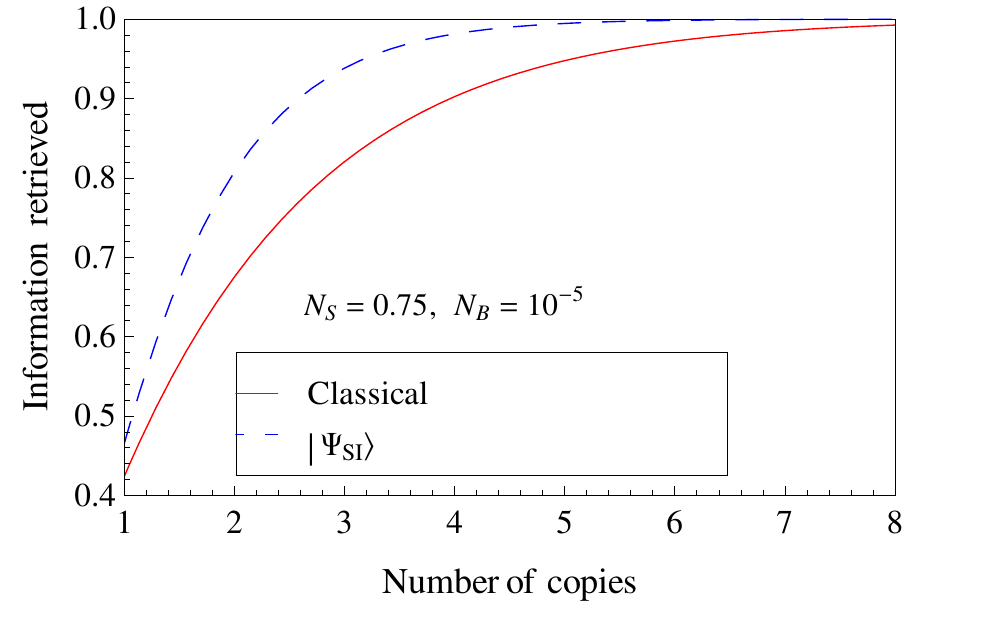}
\caption{(Color online) Maximum information   $J_{{\rm max},{\cal C}}$ retrieved from digital memory, when  classical light is used and minimum information $J_{{\rm min},{\cal Q}}$ decoded by employing quantum entangled state (\ref{newstate}) of light as a function of number of copies, M for the entangled state  for $N_S=0.75$ and $N_B=10^{-5}.$ }
\end{figure}

\section{Conclusion} 

In conclusion, following the {\em quantum reading} scheme of digital memory motivated in Ref.~\cite{pirandola}, we have explored a model of memory consisting of  perfectly reflecting mirror (reflectivity $r_1=1$), encoding bit value 1, and a completely transmitting cell of reflectivity $r_0=0$ (with stray thermal photons hitting the upper side of the memory recieved by the detector) encoding bit value 0.  This simple scheme is useful to evaluate the error rate affecting the readout of memory with  pure  entangled states of light.  We have explored the quantum reading efficiency of (i) ${\cal M}\& \,{\cal M}$ class  of path entangled  photons~\cite{Dowling} and (ii)  entangled states of light produced by combining a single photon with a  coherent state in a 50:50 beam-splitter~\cite{Gisin} by comparing them with any classical transmitter of light for fixed signal profiles, under  {\em local energy constraint}~\cite{braunstein}. It is identified that  even a single copy of faint non-Gaussian entangled light (with average signal intensity $N_S\approx 1)$ can offer improvement (with a maximum of $30\%$ information gain) in the readout of binary memory than any classical light. The results obtained here agree in general (when confined to specific values of reflectivities $r_1=1,\ r_0=0$) with the identifications of Ref.~\cite{braunstein}, where quantum-classical comparison is done by employing entangled EPR correlated light.  The improved reading efficiency recognised with  different sources of entangled light over classical light in the low signal intensity regime has implications towards  short read-out time (or high data transfer rate), ability for dense storage~\cite{pirandola, braunstein}. Consequently,  enhancement of readout in the faint light limit  -- established with different sources of quantum light -- strengthens technological possibilities of decoding  photo degradable organic memories.


\begin{thebibliography}{1}
\bibitem{note} It is convenient to express the family of ${\cal M}\&{\cal M}$ states as $\vert m::m'\rangle \equiv \vert m:: 2N_S-m\rangle,\ \ N_s<m\leq 2N_S$ in terms of the average signal intensity $N_S=(m+m')/2.$
\bibitem{note1} $M$-copy error probability of photon number states for $r_0=0, r_1=1$ is given by~\cite{ARU} $P_{\rm err}(M, N_S, N_B)=\frac{1}{2}\left(\frac{N_B^{N_S}}{(1+N_B)^{N_S+1}} \right)^M.$
\bibitem{note2} Mixing single photon Fock state with vaccuum in a 50:50 beam splitter results in the N00N state of average signal intensity $N_S=0.5$, and hence, the scheme doesn't allow for a comparison of reading efficiency of single photon state with the entangled light obtained by combining it with the vacuum state in a beam splitter. On the other hand, mixing single photon  with classical coherent light in a beam splitter enables such a comparison, as one can tune the intensity $N_S$ of the signal mode by varying the intensity of the input coherent light.
\end{thebibliography}
\end{document}